\documentclass[proof]{WileyASNA-v1}

\articletype{Original Article}%

\received{XXX}
\revised{XXX}
\accepted{XXX}

\begin{document}

\title{Denoising medium resolution stellar spectra with neural networks}

\author[1,2,3]{Balázs Pál*}
\author[3]{László Dobos}

\authormark{PÁL \textsc{and} DOBOS}

\address[1]{\orgdiv{Department of Physics of Complex Systems}, \orgname{ELTE Eötvös Loránd University}, \orgaddress{\country{Hungary}}}
\address[2]{\orgdiv{Institute for Particle and Nuclear Physics}, \orgname{HUN-REN Wigner Research Centre for Physics}, \orgaddress{\country{Hungary}}}
\address[3]{\orgdiv{Department of Physics \& Astronomy}, \orgname{The Johns Hopkins University}, \orgaddress{\state{Maryland}, \country{USA}}}

\corres{*Balázs Pál, Department of Physics of Complex Systems, Eötvös Loránd University, Pázmány Péter avenue 1/A, Budapest, Hungary. \email{pal.balazs@ttk.elte.hu}}

\abstract{We trained denoiser autoencoding neural networks on medium resolution simulated optical spectra of late-type stars to demonstrate that the reconstruction of the original flux is possible at a typical relative error of a fraction of a percent down to a typical signal-to-noise ratio of $10$ per pixel. We show that relatively simple networks are capable of learning the characteristics of stellar spectra while still flexible enough to adapt to different values of extinction and fluxing imperfections that modifies the overall shape of the continuum, as well as to different values of Doppler shift. Denoised spectra can be used to find initial values for traditional stellar template fitting algorithms and -- since evaluation of pre-trained neural networks is significantly faster than traditional template fitting -- denoiser networks can be useful when a fast analysis of the noisy spectrum is necessary, for example during observations, between individual exposures.}


\maketitle

\section{Introduction} \label{sec:intro}

Sky surveys focusing on observations of resolved stars often have to deal with the apparent brightness of target objects covering an interval of 6-7 magnitudes. Fiber-fed spectrographs on 8-meter-class telescopes can observe thousands of targets simultaneously, with the same exposure time, resulting in a broad range of signal-to-noise ratios for the final spectra. Often, the low surface density of bright stellar targets, combined with the limitations of fiber placement and the available telescope time, balances the target lists toward fainter magnitudes. This poses a challenge in both flux calibration and physical parameter inference. Atmospheric parameters, as well as the line-of-sight velocity of stars are determined by fitting theoretical stellar spectrum templates to the observed spectra which fitting process is prone to failure due to low signal-to-noise or incorrect flux calibration.

The main motivation to our work is the demanding requirements of Galactic Archaeology (GA) which investigates the structure and formation as well as the dynamical and chemical evolution of the Milky Way, the Andromeda Galaxy and their surrounding satellite galaxies. Low and medium resolution spectroscopy is an important tool to gather this information about resolved stars and new generation instruments such as DESI \citep{2022AJ....164..207D, 2023ApJ...944....1D} and Subaru PFS \citep{2014SPIE.9147E..0TS, 2014PASJ...66R...1T} are capable of observing red-giant branch (RGB) stars as far as the Andromeda Galaxy \citep{2023ApJ...944....1D} and even main sequence turn-off stars at the distance of some Milky Way satellites. By gaining insight into the dynamics of these systems via measuring the line-of-sight velocities of individual member stars, and understanding the chemical evolution by analyzing elemental abundances, we can learn about the small scale characteristics of the dark matter and its early time interaction with ordinary matter \citep{2014PASJ...66R...1T}.

The low brightness of the targets -- with the sample median easily as faint as $22.5$ magnitudes in the $i$-band, which translates to a typical $S/N$ of $5$ per pixel or $10$ per resolution element at medium resolution with 3~hours of exposure time -- poses a significant challenge for the analysis of these observations. Most stellar spectrum analysis methods rely on a continuum normalization step which first fits a template or analytic continuum to the observed flux and then divides the flux with the fitted continuum to isolate the spectral lines. Classical methods often struggle to find the continuum and best fit stellar template to noisy observations which can lead to non-converging optimizations and strongly biased parameter estimates. 

In this paper, we investigate if deep neural networks, more precisely denoising autoencoders (DAEs), are capable of recovering the underlying theoretical synthetic models from simulated medium resolution optical spectra with different levels of noise. We train and evaluate our DAE on a training set consisting of a $100{,}000$ simulated stellar spectra of varying atmospheric parameters and $S/N$. Once neural networks are trained, they can be evaluated on an observation in a few milliseconds, hence denoising autoencoders can act as a fast, on-the-fly algorithm to find the approximate best-fit synthetic spectrum and supplement more traditional algorithms to find good initial values for iterative optimization algorithms, hence reduce computation time.

\subsection{Denoising autoencoders}

Autoencoders \citep{doi.org/10.1002/aic.690370209} are neural networks that have the same, usually large number of inputs and outputs that are connected with hidden layers that have an hourglass shape with a hidden middle layer of only a small number of nodes, referred to as the \textit{bottleneck layer}. During training an autoencoder, the requirements is that it reproduces its input on the output layer as close as possible \citep{doi:10.1126/science.1127647}. The first half of the autoencoder, between the inputs and the middle layer is referred to as the \textit{encoder} and has a narrowing shape where each layer is followed by a layer with a smaller number of neurons. This first half of the network is responsible for encoding the input into as few numbers as the number of neurons in the bottleneck layer. The encoder is a non-linear mapping between the high-dimensional input and a low dimensional \textit{latent space}. The output of the bottleneck layer is a lossy compression of the input, where the compression depends on the information content of the input, the network architecture and how well the network is trained. The second half of the network, the \textit{decoder}, learns how to reconstruct the original input from the compressed representation of the bottleneck layer and it usually consists of layers with an increasing number neurons toward the output. 

Denoising autoencoders are very similar in concept to regular autoencoders (AEs) but are trained differently and can show differences in architecture \citep{Vincent2008ExtractingAC}. In case of a denoising network, the low information flux through the bottleneck layer is primarily responsible for the denoising capabilities. 

The weights of the autoencoders are trained by requiring that the network copies the input data to its output as accurately as possible. In contrast, denoising autoencoders are trained by requiring that the network produces a noiseless, or at least less noisy output from a noisy input. Since the primary goal of DAEs is to eliminate noise as much as possible, rather than encoding the input into a compressed representation as traditional AEs, the network architecture of DAEs does not necessary follow the hourglass shape and the bottleneck layers usually contains more neurons as in case of AEs.

\section{Simulating stellar spectroscopic observations} \label{sec:synthesis}

To train the denoising autoencoder, we generated a large ensemble of simulated spectra based on the BOSZ synthetic spectrum grid \citep{2017AJ....153..234B}. The observational parameters were chosen to match the typical environmental conditions at the Subaru Telescope at the Mauna Kea Observatory on dark and gray nights, as summarized in Tab.~\ref{tab:observation}. Each of the parameters is sampled uniformly within the specified intervals.

\begin{center}
    \begin{table}[t]%
        \centering
        \tabcolsep=0pt%
        \begin{tabular*}{20pc}{@{\extracolsep\fill}lcc@{\extracolsep\fill}}
            \toprule 
            \textbf{parameter}   & \textbf{range}        \\ \midrule
            seeing               & $0.5$ - $1.5''$       \\
            target zenith angle  & $0$ - $45^\circ$      \\
            field angle          & $0.0$ - $0.65^\circ$  \\
            moon zenith angle    & $30$ - $90^\circ$     \\
            moon target angle    & $60$ - $180^\circ$    \\
            moon phase           & $0.0$-$0.25$          \\
            exposure time        & $15$~min              \\
            exposure count       & $20$                  \\
            \bottomrule
        \end{tabular*}
        \caption{Summary of the observational parameters and parameter ranges that were sampled uniformly to generate the simulated spectra. These parameters are very similar to the typical conditions at at Mauna Kea Observatory, except for the seeing which tends to be $0.8''$ or higher. The exposure time was set to 5~hours, with 20 exposures and detector read-outs, 15~minutes each.\label{tab:observation}}%
    \end{table}
\end{center}

We simulated the spectra as if they were observed with an instrument very similar to the $R \approx 5000$ medium resolution mode of the Subaru Prime Focus Spectrograph \citep{2016SPIE.9908E..1MT}. The parameters of the spectrograph are listed in Tab.~\ref{tab:instrument}. The instrument was specifically designed for Galactic Archeology and its wavelength range was chosen to cover lines that carry the most information about line-of-sight velocity and elemental abundances in a limited wavelength range.

\begin{center}
\begin{table}[t]%
    \centering
    \tabcolsep=0pt%
    \begin{tabular*}{20pc}{@{\extracolsep\fill}lcc@{\extracolsep\fill}}
        \toprule 
        \textbf{parameter}  & \textbf{MR mode}  \\ \midrule
        $\lambda$ coverage  & $710$ to $885$~nm \\
        pixel dispersion    & $0.4$~\AA         \\
        spectral resolution & $1.6$~\AA         \\
        velocity resolution & 60~km~s$^{-1}$    \\
        resolving power     & 5000              \\
        \bottomrule
    \end{tabular*}
    \caption{Summary of the spectrograph parameters similar to the medium resolution of the Prime Focus Spectrograph of the Subaru Telescope.\label{tab:instrument}}%
\end{table}
\end{center}

The ranges of fundamental stellar parameters were chosen to cover stellar types from M to G, which differ from the typical targets of Galactic Archeology, which prefers the stellar types G and F. We chose the lower temperature range, because of the higher variance of the absorption features, which makes the denoising task more challenging and exercises the capabilities of the neural network to learn the spectral features.

Since the parameters are uniformly and independently drawn from continuous intervals, the simulated spectra include many non-physical models which should have no significant impact on the performance of the neural networks trained with them. The input parameters of the simulations are summarized in Tab.~\ref{tab:stellar}. We used these values to generate both, the training set and the validation set for optimizing the weights of the denoising autoencoder, as detailed in Sec.~\ref{ssec:data}.

\begin{center}
    \begin{table}[t]%
        \centering
        \tabcolsep=0pt%
        \begin{tabular*}{20pc}{@{\extracolsep\fill}lcc@{\extracolsep\fill}}
            \toprule
            \textbf{parameter}         &                  & \textbf{range}      \\ \midrule
            effective temperature      & $T_\mathrm{eff}$ & $3500$ - $5000$~K              \\
            surface gravity            & $\log g$         & $0.0$ - $5.0$                  \\
            metallicity                & $[\mathrm{Fe}/\mathrm{H}]$ & $-2.5$ - $0.75$~dex  \\
            $\alpha$ element abundance & $[\alpha/\mathrm{H}]$      & $-0.25$ - $0.5$      \\
            Doppler shift              & $z$              & $-0.003$ - $0.003$             \\
            magnitude in the $g$-band  & $m$              & $19.0$ - $22.5$~mag            \\
            extinction                 & $E(B-V)$         & $0.0$ - $0.1$~mag              \\
            \bottomrule
        \end{tabular*}
        \caption{Summary of the of the fundamental stellar atmospheric parameters and other simulation input parameters. The intervals are sampled uniformly in their respective ranges to generate the simulated spectra. The fundamental parameters cover the range of interest of Galactic Archeological observations but also include non-physical models that we did not exclude.\label{tab:stellar}}%
    \end{table}
\end{center}

The spectrum simulation process can be separated into two main parts. The first part is the interpolation of the synthetic stellar grid to calculate a model spectrum at a randomly chosen set of atmospheric parameters and the second part is the simulation of all observational and instrumental phenomena that affect the light of the star between leaving its atmosphere and getting detected by the instrument. The output of the simulation is a noiseless spectrum with the specific flux expressed in physical units, as well the variance of the flux for each spectral pixel. The variance vector is later used to generate realizations of the noise during the training process of the neural network. 

In addition to the precomputed observational effect, we apply further transformations to the simulated spectra during training of the neural network which serve the purpose of dataset augmentation. For instance, we generate new realizations of the photon noise on-the-fly, at each training epoch as well as introduce a wavelength-dependent systematic flux calibration error to mimic the effects of improper fluxing. These will be detailed in Sec.~\ref{ssec:augment}.

The observational and instrumental effects are modelled according to the Exposure Time Calculator (ETC) of the Subaru PFS. The Subaru PFS ETC is an adaptation of the original WFIRST ETC code, as detailed in \cite{2012arXiv1204.5151H}. In order to be able to generate hundreds of thousand of spectra, we modified the ETC code to output separately (i) the atmospheric and instrumental transmission functions, (ii) the individual noise terms originating from the night sky, as well as (iii) the instrumental line spread function (LSF) as a function of wavelength. In turn, the LSF, at each wavelength pixel, was fitted with a Gaussian function with a wavelength-dependent width so that it can be evaluated at any wavelength and spectral resolution, consistently with the wavelength grid of the input synthetic spectra.

We ran the modified ETC on a grid of observational parameters and ingested the outputs into data arrays. Our parallel Python implementation of the observation simulator interpolates these precomputed grids to calculate and sum up the noise terms that contribute to the flux variance.

Below, we discuss the simulation steps in order, and explain further processing that happens during the training process only.

\subsection{Spectrum interpolation} \label{ssec:interpolation}

We start the simulations by interpolating a regular grid of fluxed synthetic stellar spectra which are higher resolution than the instrument in order to be able to confidently evaluate the sensitivity of the denoiser to sub-pixel Doppler shifts. We chose to use the BOSZ synthetic stellar grid~\citep{2017AJ....153..234B} of ATLAS9 models \citep{1979ApJS...40....1K, 2012AJ....144..120M}. The BOSZ grid is available at a broad range of resolutions and its underlying model and line lists are considered reliable in the parameter ranges we are interested in. We interpolate the flux using one-dimensional cubic splines in the direction of a single parameter only, chosen randomly from the four fundamental parameters $[\mathrm{Fe}/{H}]$, $T_\mathrm{eff}$, $\log g$ and $[\alpha/\mathrm{Fe}]$, meanwhile the rest of the parameters are kept confined to grid values.

When we later evaluate the performance of the denoising autoencoder, as a function of the each of fundamental parameters, we will only consider the models that are interpolated in the direction of the particular parameter.

\subsection{Observational and instrumental effects} \label{ssec:effects}

After interpolating the synthetic stellar grid, we apply each observational effect in the specific order as follows.

\begin{enumerate}
    \item Draw a value for the Doppler shift from a uniform distribution between $-0.003 \leq z \leq 0.003$, and shift the wavelength of the interpolated model accordingly, without rebinning or resampling the spectrum.
    \item Draw a value for the extinction randomly from a uniform distribution between $0 \leq E(B-V) \leq 0.1$ and apply the mean extinction curve of the diffuse dust in the Milky Way, as determined by \cite{1989ApJ...345..245C}, to the Doppler-shifted spectrum.
    \item Convolve the spectrum, still at its original high resolution, with a line spread function similar to the LSF of full optical system of Subaru PSF, including the pixelization of the detector.
    \item Interpolate the spectrum to the wavelengths of the detector pixels. In our case, the number of detector pixels is $4096$.
    \item Given a randomly drawn apparent magnitude, calculate the flux in physical units as well as the photon count from the source and the different components of the night sky including the continuum, sky lines and reflected sunlight from the Moon.
    \item Calculate the flux variance in each detector pixel from the photon counts and detector characteristics as if all exposures were stacked together into a single spectrum.
    \item Normalize the flux to have a median value of $0.5$ in the wavelength range of $6500$-$9500~\Angstrom$.
\end{enumerate}

Finally, we measure the signal-to-noise as the median of the ratio of the flux to the square root of its variance in each pixel over the entire wavelength coverage of the instrument. Instead of using $S/N$ as the input parameter, we normalize the synthetic flux of the model spectrum to a magnitude drawn from a uniform distribution that covers the typical magnitude range of potential targets. Although this method will not result in uniform sampling in $S/N$, it will make sure that the flux error from sky subtraction, scattered light and the noise contribution of the detector are taken into at account at the right ratio. Rescaling the noise level at a later step to a given $S/N$ would change the ratio of the noise components.

Real observations contain many pixels that need to be masked due to various reasons such as cosmic rays, bright sky lines, low detector efficiency etc. In this paper, we do not mask pixels of the simulated spectra and do not train the denoising autoencoder to work with masked inputs.

The final normalization step is necessary to transform the flux values of the simulated spectra into the $[0; 1]$ interval, which our deep neural network with a sigmoid activation function at its output can handle.

We assume that the multiple exposures of the same target were taken at the same observation conditions, hence we only run the simulation once and assume perfect stacking, with perfectly matching wavelengths and LSF when calculating the variance of the flux.

\begin{figure}
    \includegraphics[width=\columnwidth]{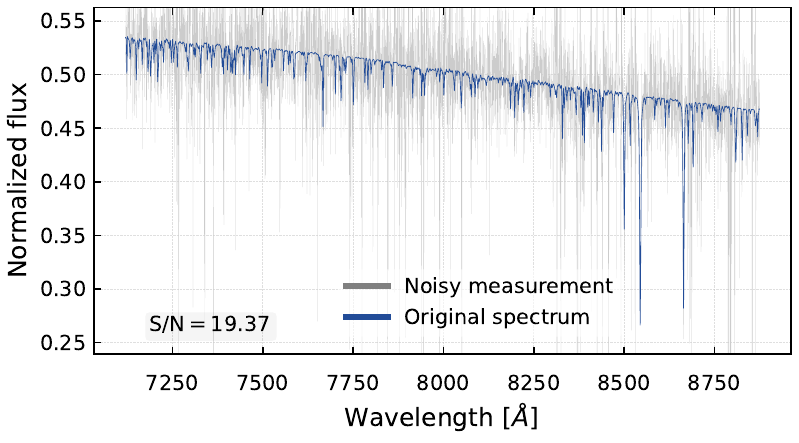}
    \caption{An example of a noiseless simulated stellar spectrum observation and its noisy counterpart. A noise component was added to the original spectrum as described in Sec.~\ref{ssec:augment}. The spectrum is rescaled by setting the median of the flux in the $6500$-$9500~\Angstrom$ wavelength range to $0.5$. The $S/N$ value indicated in the bottom left corner is the median pixel-wise signal-to-noise ratio over all pixels.}
    \label{fig:noise}
\end{figure}

Fig.~\ref{fig:noise} shows a typical simulated spectrum of a star with parameters $[\mathrm{Fe}/\mathrm{H}]=-1.5$, $T_\mathrm{eff} = 4183$~K, $\log g = 2.0$, $[\alpha/\mathrm{Fe}] = 0.0$, with and without a random realization of the noise added to the flux. The purpose of the denoising autoencoder is to recover the noiseless spectrum (blue curve) from the noisy output (gray curve).

In addition to the observational effects detailed above, during the training process, as part of the data set augmentation, we multiply each spectrum with a different, randomly generated function that varies slowly with wavelength. This was done to simulate the effects of incorrect flux calibration. The function is generated using a random walk process, filtered with a low-pass filter, resulting in a higher-order function than typical polynomial corrections used for flux calibration. Although the flux calibration effects are most often systematic, training the network with a randomized flux calibration error will make it resilient against both random and systematic calibration errors. The maximum amplitude of the flux calibration error was set to be $2$~per~cent, which is a typical value \citep{2014AJ....147..127B}.

\section{Denoiser implementation} \label{sec:training}

To realize a denoising autoencoder, we trained a relatively shallow, hourglass-shaped, fully connected network with rectified linear unit activation functions between the hidden layers, and a sigmoid function before the output using stochastic gradient descent with no momentum. The optimization of the weights was based on a training and a validation data set with $100{,}000$ simulated spectra in each. In addition, we augmented the data sets with a different realization of the flux noise at each training step. During training, the gradient at each step was evaluated by setting the network input to the noisy spectrum, while the output was required to be the noiseless spectrum.

The network architecture, data set augmentation and training process are described in detail in the following sub-sections.

\subsection{Neural network architecture} \label{ssec:arch}

\begin{figure*}
  \centering
  \includegraphics[width=\linewidth]{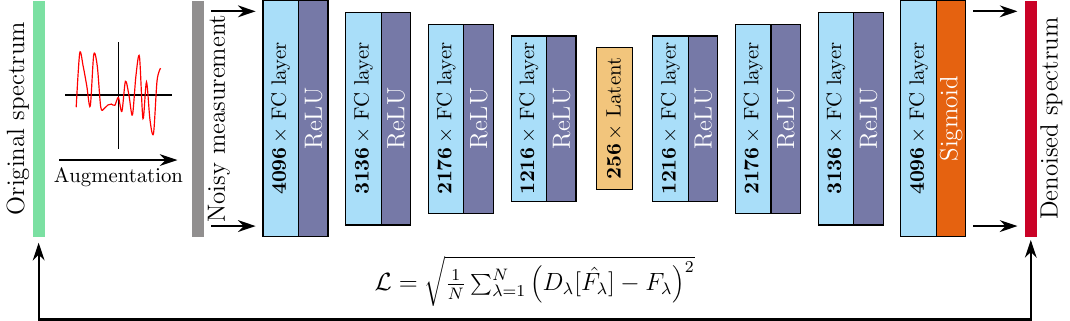}
  \caption{Schematic overview of our denoising autoencoder and the training pipeline including the augmentation steps. The noiseless input spectra, labeled as ''original spectrum'', were simulated as described in Sec.~\ref{sec:synthesis}. During the training process, these spectra are fed into an augmenter, described in Sec.~\ref{ssec:augment}, resulting in the ``noisy measurement''. Following this, the noisy observations are processed by the DAE network, which is trained to output the denoised spectra, labeled here as ``denoised spectrum''. During the training, a stochastic gradient descent optimizer is used to find the network weights that minimize the RMSE loss calculated from the ``denoised spectrum'' and the ``original spectrum''.}
  \label{fig:architecture}
\end{figure*}

The architecture of the denoising autoencoder we used in this study is illustrated in Fig.~\ref{fig:architecture}. Both, the encoder and the decoder side of the network consist of $4$-$4$ fully-connected layers, with an input and output size of $4096$, the number of wavelength pixels of the simulated spectroscopic observations. The number of neurons in each hidden layer of the encoder decreases linearly, while on the decoder side, the number of neurons increases the same way. Each fully-connected layer is followed by a bias vector with the same size as the output of that layer. After each layer, except prior to the output, a rectifier linear unit (ReLU) activation was used. Before the output, a smoother sigmoid activation was inserted instead, to produce the final output.

Before training, the weights, as well as the bias vector of the fully-connected layers are initialized using the Xavier initialization scheme with values drawn from the uniform distribution $\mathcal{U}(-\sqrt{k}, \sqrt{k})$, where $k = 6 / (\mathrm{inputs} + \mathrm{outputs})$, a function of the number of inputs and outputs of the layer. The size of the bottleneck layer was set to a relatively large number of $256$ neurons.

In summary, all layers in our autoencoder network follow the conventional fully connected feed-forward neural network architecture as described by
\begin{equation}
    \mathbf{Z} = \sigma \left( \mathbf{W} \mathbf{x} + \mathbf{b} \right),
\end{equation}
where $\mathbf{x}$ is the input vector, $\mathbf{W}$ is the weight matrix, $\mathbf{b}$ is the bias vector, $\sigma$ is the activation function, with $\mathbf{Z}$ being the output vector of the layer.

To implement the network, we used the PyTorch deep learning framework \citep{NEURIPS2019_9015}.

\subsection{Training and validation set properties} \label{ssec:data}

Using the stellar parameter distributions described in Sec.~\ref{sec:synthesis} and summarized in Tab.~\ref{tab:stellar}, we generated separate data sets for training and validation, each with $100,000$ simulated spectra. As we have used entirely synthetic data sets, we did not use a third, dedicated test set for the evaluation of the trained network, instead we used the validation set for this purpose.

As mentioned in Sec.~\ref{ssec:effects}, the spectra, and the corresponding flux variance array were normalized, so that flux values cover the $[0, 1]$ interval, thus no further input or output normalization happens during the training of the network. However, we point out that the denoiser is unable to produce an absolutely noiseless spectrum. When some remaining noise is present that extends beyond the $[0, 1]$ interval (i.e., part of the predicted spectrum is outside this interval), the output values are clipped to this range. This is because the output of the network is bounded by the sigmoid function at the end of the network, which produces values in the $[0, 1]$ interval. Consequently, this results in a slightly non-Gaussian residual noise in the output.

\subsection{Data augmentation} \label{ssec:augment}
In addition to generating a large number of simulated spectra, we further augment the number of training samples by on-the-fly transformations of the input and the output during the training process. The steps of the augmentation process are as of follows.

\begin{enumerate}
    \item Generate a random realization of the flux calibration error and apply it to both the input and output.
    \item Generate a random realization of the observational noise and add it to the input spectrum. The output is kept noiseless.
    \item Generate random values for the scale and offset and apply to both the input and the output.
\end{enumerate}

At each training epoch, we generate a new noise realization for every spectrum based on the flux variance vector. Using new realizations of the noise at each training epoch is a very efficient way of preventing overfitting the training set and, since noise in inherent to spectroscopic observations, it is also a very natural method of data augmentation.

In addition to adding noise to the spectra in each epoch, we generate a new random realization of the flux calibration error, as explained in Sec.~\ref{ssec:effects}, that we apply to both, the noisy input and the noiseless output spectrum. This is done to increase the robustness of the network against systematics in the flux calibration.

Similarly, in each epoch, we further augment the data sets by applying a small, random offset (additive term) and scale (multiplicative term), both drawn from the normal distributions $\mathcal{N}(0, 0.01)$ and $\mathcal{N}(1, 1.01)$, respectively.

\subsection{Loss function} \label{ssec:loss}

Let us denote the simulated noiseless flux with $F_\lambda$ and the noisy spectrum with $\hat F_\lambda$, whereas the denoised spectrum with $D_\lambda [ \hat F_\lambda ]$ where $D$ is the multivalued function describing the denoiser and $\lambda$ indexes the spectral pixels.
In the training loop, we minimize the root-mean-square error (RMSE) with the usual definition of
\begin{equation}
    \mathcal{L}
    =
    \sqrt{\frac{1}{N} \sum_{\lambda=1}^{N} \left( D_\lambda [ \hat{F_\lambda} ] - F_{\lambda} \right)^{2}},
\end{equation}
where $N$ is the number of spectral pixels.

We decided to use RMSE as the loss function for two reasons. The first one is its interpretability in the comparison of noisy and noiseless stellar spectra. In the ideal case, when the denoiser efficiently reduces the noise of the input, without introducing any artifacts, the RMS error measures the amount of residual noise. The second reason is its sensitivity to larger differences between the actual output and the ideal one but it is easier to optimize using stochastic gradient descent than, for instance, the $L_0$ norm, which is only sensitive to the spectral pixel with the largest error.

We expected that the largest errors would appear at the cores of the deepest absorption lines, as those are the features with the largest variance in the training sample. On the other hand, we aimed for a noise reduction that is more or less uniform over the entire wavelength coverage and also takes shallower absorption features into account with approximately equal weight. The mean absolute error (MAE) turned out to be too unsensitive at the strong spectral lines.

\subsection{Training}

We optimized the weights of the network with the stochastic gradient descent (SGD) algorithm \citep{10.1214/aoms/1177729586} using a custom learning rate schedule. SGD is an iterative method that aims to minimize the loss function by updating the weights of a model based on the gradient of the loss with respect to the weights.

Notably, our implementation of SGD did not utilize Nesterov momentum, which is often used to accelerate convergence in relevant directions and dampen oscillations \citep{Nesterov1983, pmlr-v28-sutskever13}. We made this choice to maintain simplicity and control of the learning process that we wanted to drive only with an adaptive learning rate scheduler.

Due to the large number of outputs and the sigmoid activation function before the network output the training experienced the problem of vanishing gradients. With a training batch size of $250$, the typical values of the gradient for the first layer happened to be as small as $10^{-6}$-$10^{-7}$ on average. To mitigate this problem without batch normalization, we trained our model with no L2 regularization (weight decay) and set the learning rate to an unusually high starting value of $100$.
While this is generally discouraged due to the instabilities it can introduce into the loss function, we found that this unconventional technique greatly helped our model to converge.

\begin{figure}
    \includegraphics[width=\columnwidth]{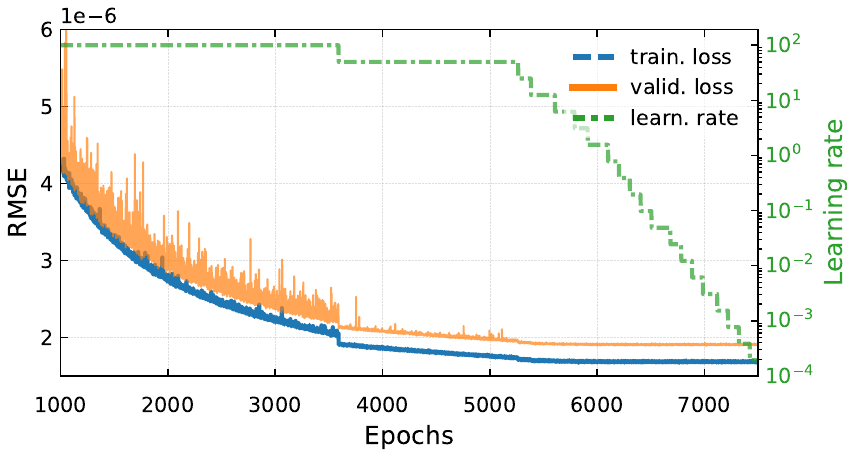}
    \caption{Training and validation loss during a $7{,}500$ epoch long training process, starting from the $1{,}000$th epoch. All loss values shown here are the epoch averages for each spectrum of the training (validation) data set. The irregularly fluctuating line of the validation loss sits on top of the smoother curve of the training loss, indicating that our model was neither under nor over-fitted. The value of the adaptive learning rate is also shown here in green. The learning rate is a series of step function according to how the adaptive learning rate scheduler decreases its value during the training process. \textbf{Note:} Cutting the first $1{,}000$ epochs from the plot served the purpose of better visualization of the loss values. The initial epochs were characterized by an extremely steep loss decrease and including them would have made the plot less readable.}
    \label{fig:loss}
\end{figure}

During training, we adaptively reduced the learning rate when the model stopped improving for a predefined number of epochs, as measured from the validation loss. After some patience period, the learning rate was halved as seen in Fig.~\ref{fig:loss}. This approach allows the model to escape local minima and to fine-tune its parameters as the training progresses, ultimately leading to a more robust and generalized solution. In our case, the value of the patience parameter was set to $150$ epochs.

\section{Results} \label{sec:results}

In this section we analyze the results obtained by training a denoising autoencoder neural network as described in Sec.~\ref{sec:training}.

\subsection{Measures of goodness} \label{ssec:errors}

We measure the absolute and relative error after denoising by comparing the denoised spectrum $D_\lambda[\hat F_\lambda]$ to the original simulated spectrum without any additional noise $F_\lambda$, where $\hat F_\lambda$ is the noisy spectrum. The definition of the absolute error for each pixel is
\begin{equation} \label{eq:absolute_error}
    | E_\lambda |
    =
    \left| D_\lambda[\hat F_\lambda] - F_\lambda \right|,
\end{equation}
whereas the relative error is taken to be
\begin{equation} \label{eq:relative_error}
    | R_\lambda |
    =
    \frac{\left| D_\lambda[\hat F_\lambda] - F_\lambda \right|}{F_\lambda}.
\end{equation}
When we characterize the denoising performace with a single number, we simply take the mean of $| E_\lambda |$ or $| R_\lambda |$ over every wavelength pixel.

To provide information on the bias, we also report error metrics without taking the absolute value of the difference $D_\lambda[\hat F_\lambda] - F_\lambda$. We denote these metrics without the absolute value as $E_\lambda$ and $R_\lambda$.

\subsection{Denoising performance} \label{ssec:denoising}

\begin{figure}
    \includegraphics[width=\columnwidth]{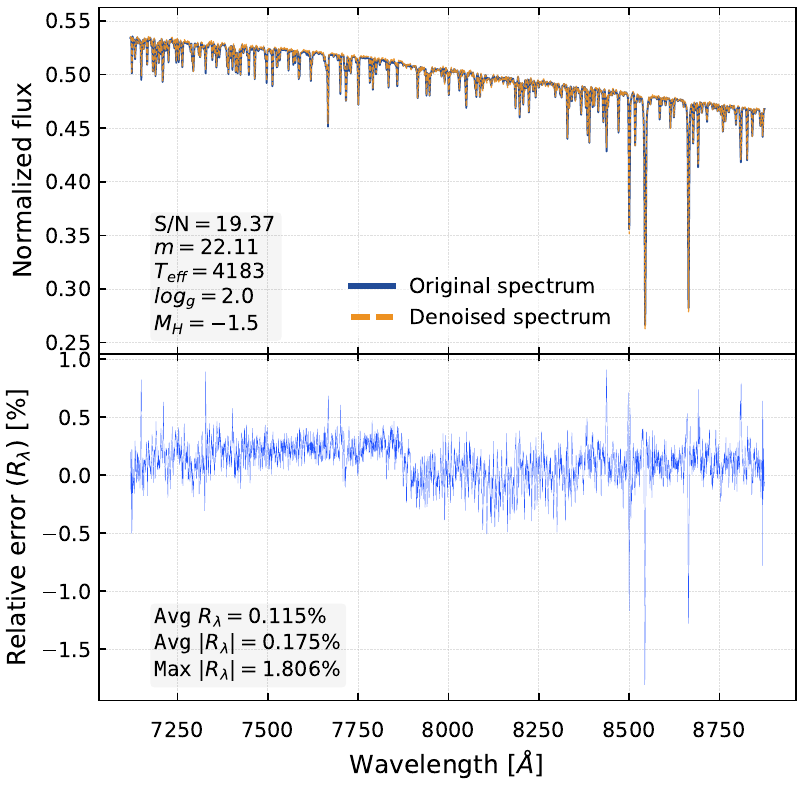}
    \caption{
    \textbf{Top panel:} Comparison of the original noiseless simulated stellar spectrum (solid blue line) and the denoised spectrum (dashed orange line) using our denoising autoencoder. Note that the two lines that represent the original and the denoised spectrum are almost entirely overlap, indicating high accuracy of our model.
    \textbf{Bottom panel:} The relative error calculated as the fraction of pixel-wise residual noise and the original noiseless flux. The mean and maximum of the relative error, as well as the Kullback-Leibler divergence of the original noiseless spectrum and the denoised spectrum are also indicated. Note, that the peaks of the relative error are aligned with the most prominent absorption features.}
    \label{fig:prediction}
\end{figure}

\begin{figure}
    \includegraphics[width=\columnwidth]{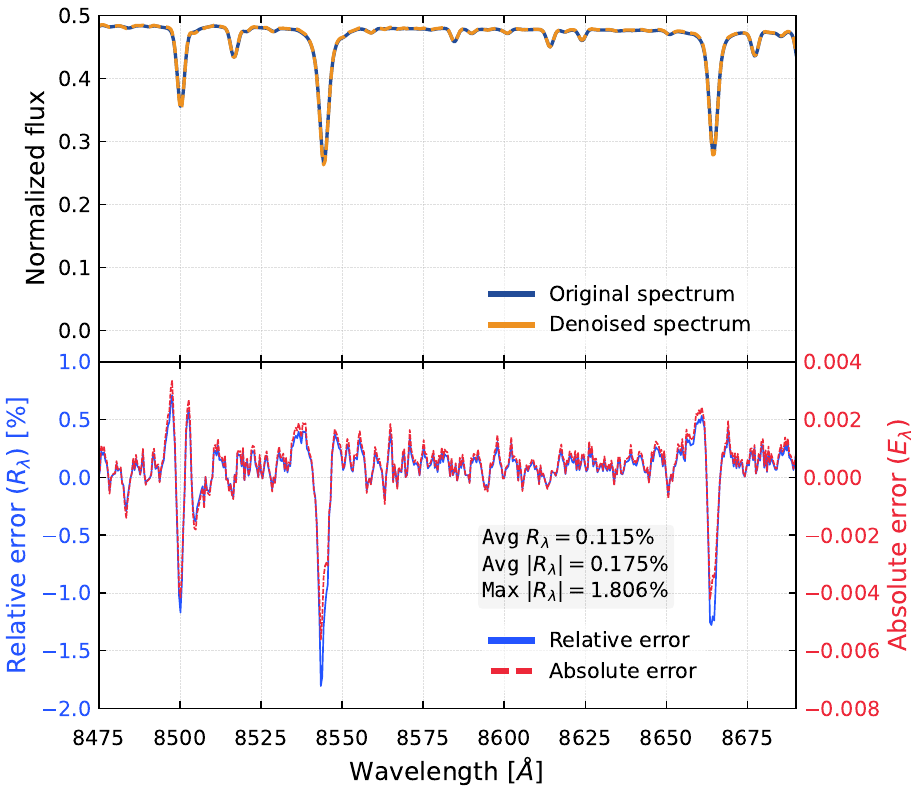}
    \caption{A zoomed-in view of the calcium triplet for the same spectrum as seen in Fig.~\ref{fig:prediction}. In addition to the relative error, the bottom panel also shows the pixel-wise absolute errors calculated as the absolute value of the difference between the denoised spectrum and the noiseless original spectrum.}
    \label{fig:prediction-zoom}
\end{figure}

As an example, in Fig.~\ref{fig:prediction}~and~\ref{fig:prediction-zoom}, we plot the spectrum of a typical G giant and its reconstruction from a noisy simulated observation, as well as the residual and the relative error calculated as the ratio of the residual and the noiseless flux in each spectral pixel. As one would expect, the error is larger in the cores of the strong absorption lines, such as the lines of the calcium triplet because stellar spectra show a larger variance in the lines than in the continuum.

\begin{figure}
    \includegraphics[width=\linewidth]{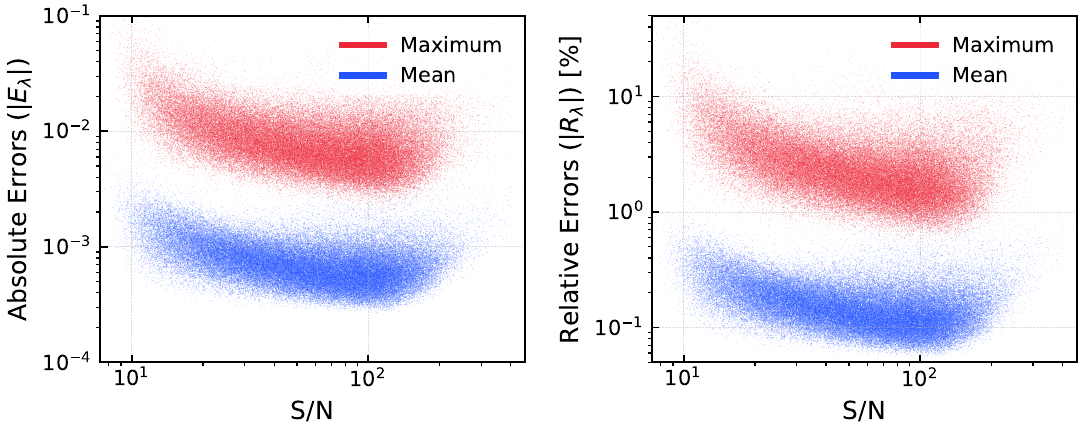}
    \caption{Pixel-wise mean and maximum of the absolute and relative errors calculated from the residual noise of the denoised spectra for the entire validation data set as a function of median $S/N$. See the text for discussion.}
    \label{fig:snr-errors}
\end{figure}

\begin{figure}
  \includegraphics[width=\linewidth]{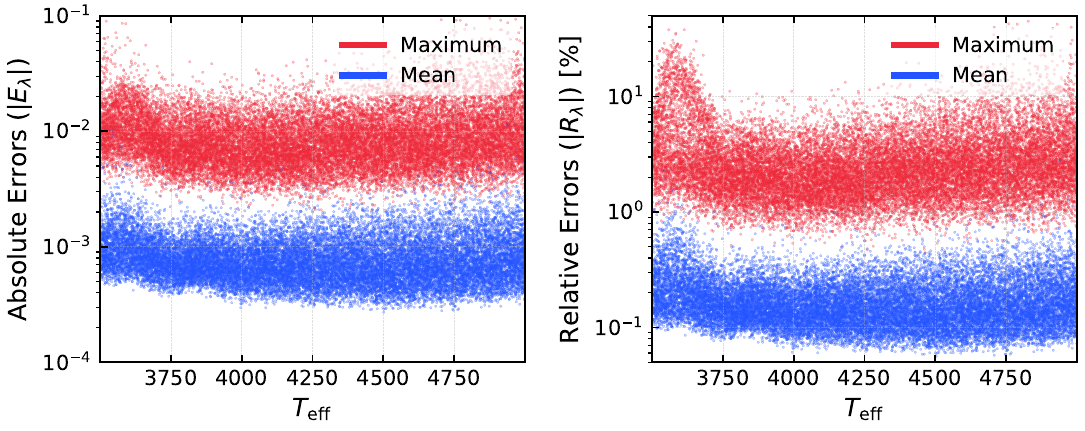}
  \caption{Pixel-wise mean and maximum of the absolute and relative errors calculated from the residual noise of the denoised spectra for the entire validation data set as a function of $T_{\mathrm{eff}}$. See the text for discussion.}
  \label{fig:teff-errors}  
\end{figure}

\begin{figure}
  \includegraphics[width=\linewidth]{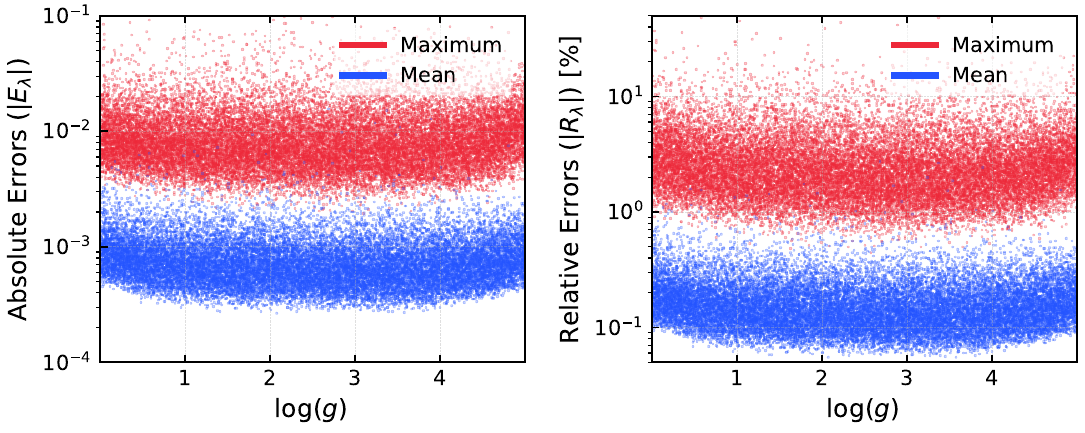}
  \caption{Pixel-wise mean and maximum of the absolute and relative errors calculated from the residual noise of the denoised spectra for the entire validation data set as a function of $\log g$. See the text for discussion.}
  \label{fig:logg-errors}  
\end{figure}

\begin{figure}
  \includegraphics[width=\linewidth]{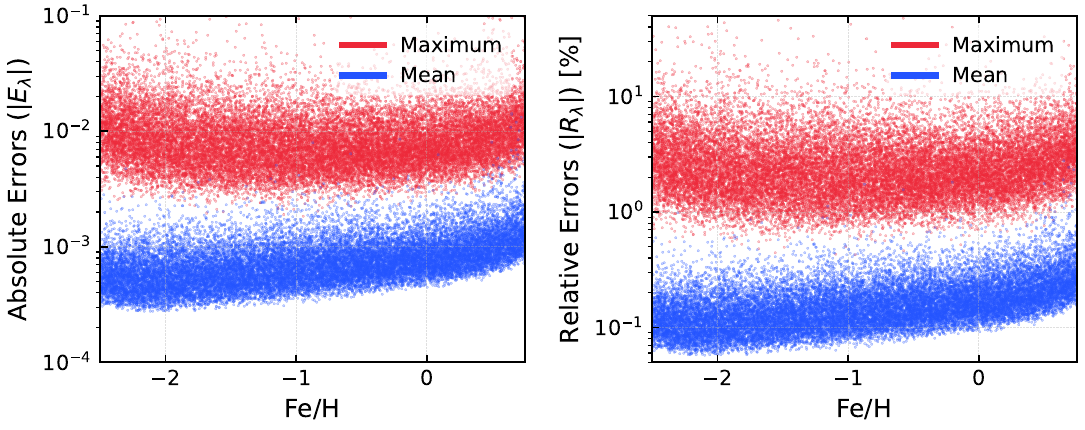}
  \caption{Pixel-wise mean and maximum of the absolute and relative errors calculated from the residual noise of the denoised spectra for the entire validation data set as a function of $[\mathrm{Fe}/\mathrm{H}]$. See the text for discussion.}
  \label{fig:mh-errors}  
\end{figure}

To get a more general view, we plot the maximum and mean of the absolute errors $| E_\lambda |$ and relative errors $| R_\lambda |$ as functions of the signal-to-noise in Fig.~\ref{fig:snr-errors} as well as functions of the fundamental stellar parameters in Fig.~\ref{fig:teff-errors}-\ref{fig:mh-errors}. We use the absolute values of the error metrics to generate data for these figures. This approach prevents the cancellation of positive and negative values in $E_\lambda$ and $R_\lambda$, thereby preserving information on the magnitude of the errors.

Unsurprisingly, as visible in Fig.~\ref{fig:snr-errors}, the performance of denoising scales with the median $S/N$ of the spectra but the means of both the absolute error and the relative error are within a decade between $10 < S/N < 120$. The errors show a significant upturn toward lower $S/N$ which limits the applicability of out network for really noisy spectra.

When looking at the residual noise as a function of $T_\mathrm{eff}$ in Fig.~\ref{fig:teff-errors}, one can observed that the amount of remaining noise depends only slightly on temperature except for $T_\mathrm{eff} < 3750$~K where the relative error shows a significantly larger spread than elsewhere. This can easily be attributed to the complexity of the optical spectra of M dwarfs due to the strong blanketing by molecular lines.

Some dependence of the residual error is visible as a function of $\log g$ in Fig.~\ref{fig:logg-errors}. The mean of the relative error is a few times larger toward the most extreme values of the validation set which is a typical result for networks trained and validated on data sets with the same parameter range coverage. The dependence of the residual noise on $[\mathrm{Fe}/\mathrm{H}]$, on the other hand, shows a more peculiar behaviour as it can be seen in Fig.~\ref{fig:mh-errors}. Higher metallicity stellar spectra are reconstructed by the denoiser network with larger residual noise, very likely due to the presence of more and stronger absorption lines. Interestingly, the mean relative error scales more steeply with metellicity than the absolute error, hinting that a large number of spectral pixels is responsible for the increase in overall noise residual.

\subsection{Tests of generalizability} \label{ssec:artificial}

To assess the robustness and generalizability of our denoising autoencoder, we took noiseless simulated spectra and added new, strong, artificial absorption lines with a Voigt profile and also increased the depth of some existing absorption lines. After adding the artificial lines, we added noise to the spectra and processed them with the denoising autoencoder that was trained with realistic absorption lines only.

Our quite plausible hypothesis was that the denoising autoencoder, more specifically its decoder, works by learning how to generate synthetic spectra from the latent space (i.e. global features) as opposed to learning how to remove the noise and reconstruct the continuum and spectral lines from local features.

\begin{figure}
    \includegraphics[width=\columnwidth]{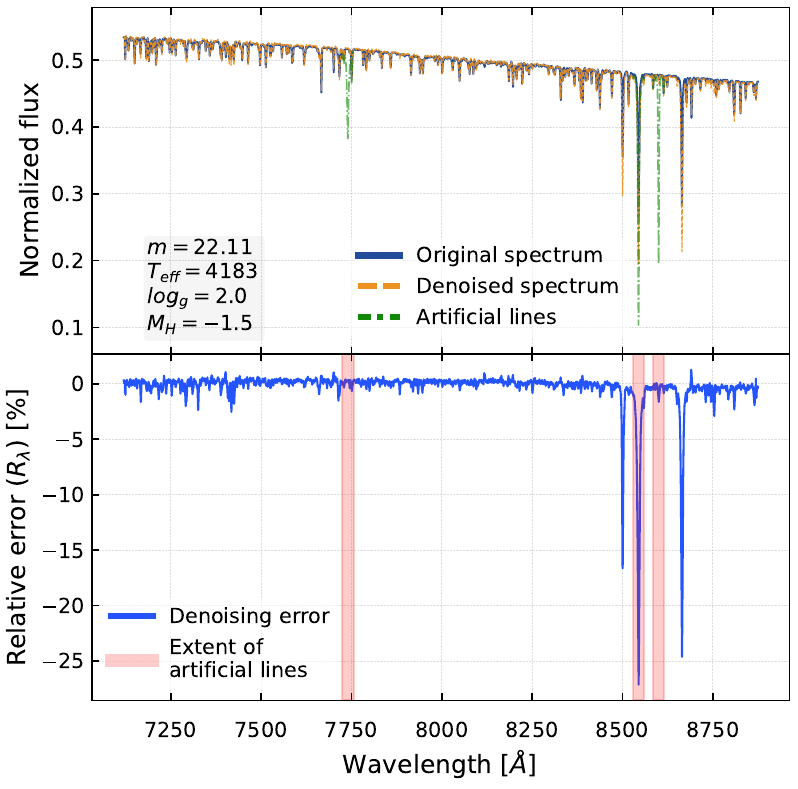}
    \caption{
    \textbf{Top panel:} Comparison of a noiseless simulated spectrum (blue line) with artificially introduced extra absorption features (green line) to its denoised version (orange line). The denoised spectrum very precisely covers the original spectrum without the artificially added absorption lines.
    \textbf{Bottom panel:} The relative error of the denoising calculated as the pixel-wise absolute difference between the original noiseless spectrum with the artificial absorption features and the denoised spectrum, normalized by the noiseless flux. The relative error clearly indicates that the denoiser is unable to reconstruct the artificially introduced features which cause the peaks in relative error. Interestingly, additional peaks appear in the relative error at absorption lines which were not artificially modified, such as the two longer wavelength lines of the calcium triplet. This can be attributed to the strong correlation between the calcium lines that the denoising network learned from the training data and how the denoiser fails to benefit from the correlation when the depth of one of the lines is artificially changed.}
    \label{fig:artificial}
\end{figure}

In Fig.~\ref{fig:artificial}, we plot the calcium triplet region of a G giant star with artificially introduced additional absorption features. As the green curve of Fig.~\ref{fig:artificial} indicates, the depth of the $8542$~\AA{} feature was increased whereas a completely new feature was introduced at $8600$~\AA{}. Fig.~\ref{fig:artificial} shows the denoised spectrum in orange which exhibits the original line depth for the $8498$~\AA{} line and completely lacks the artificially introduced extra feature.

This behaviour indicates that instead of simply performing pixel-wise denoising and returning the artificially introduced features without noise, the model learns the fundamental parameters from the simulated spectra and reconstructs the denoised spectrum from those, instead of ``blind'' denoising of the input.

\begin{figure}
    \includegraphics[width=\columnwidth]{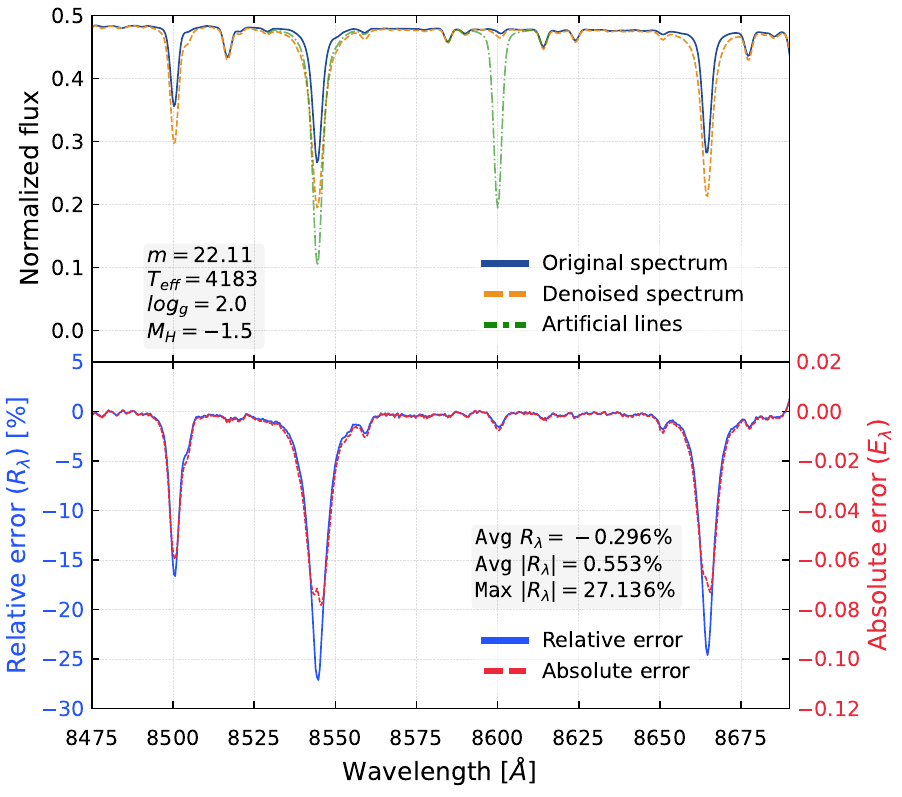}
    \caption{A zoomed-in view of the calcium triplet of the same spectra as in Fig.~\ref{fig:artificial}. The bottom panel also includes a comparison of the pixel-wise relative and absolute errors of denoising.}
    \label{fig:artificial-zoom}
\end{figure}

\section{Conclusion} \label{sec:conclusion}


We found that even simple neural networks, specifically denoising autoencoders (DAEs) consisting of fully connected layers with a bottleneck are capable of learning the characteristics of simulated stellar spectrum observations, We demonstrated that this capability enables the autoencoders to efficiently denoise simulated spectroscopic observation. For a wide range of fundamental stellar parameters, denoising can be performed at a sub-per-cent accuracy at $S/N \approx 10$.

By introducing artificial absorption features into the simulated spectra, we were able to prove that the denoiser works by representing physical spectra in a low-dimensional latent space and reconstructs the output based on that instead of removing the noise by some kind of effective interpolation of local features. As a consequence, noise reduction capabilities depend on how well the encoder part can map the input noisy spectra into the latent space of the network bottleneck layer, as well as how accurately the decoder can transform the latent space variables into noiseless spectra. This knowledge can help training more complex networks with larger training sets that perform similarly well on noisier simulations and real observations.



\section*{Acknowledgments}
This work is supported by the generosity of Eric and Wendy Schmidt, by recommendation of the Schmidt Futures program, the Ministry of Innovation and Technology NRDI Office grants OTKA NN 129148, the KDP-2021 program of the Ministry of Innovation and Technology from the source of the NRDI fund and by the the European Union project RRF-2.3.1-21-2022-00004 within the framework of the MILAB Artificial Intelligence National Laboratory. 

The simulation of all stellar spectrum observations used in this research, and the training of the DAE networks were completed on the \textit{Elephant} and \textit{Volta} servers at the Department of Physics \& Astronomy of the Johns Hopkins University. Analysis was carried out on the \textit{Volta} server, as well as on the \textit{Ampere} A100 GPU server of the Wigner Scientific Computing Laboratory (WSCLAB) at the HUN-REN Wigner Research Centre for Physics, Hungary.

\bibliography{./bibliography}%

\end{document}